\documentclass[usenatbib]{mn2e}
\usepackage{amssymb, amsmath}
 \title[ On the Doppler effect for light from orbiting sources in
Kerr-type metrics]{\bf On the Doppler effect for light from orbiting sources in
Kerr-type metrics}
 \author[S. ~Cisneros et. al]{S.\,~Cisneros$^{1,2}$\thanks{E-mail:cisneros@mit.edu}, G.\,~Goedecke$^{2}$, C.\,~Beetle$^{3}$, and  M.\,~Engelhardt$^{2}$\\
 $^{1}$ Department of Physics, Massachusetts Institute of Technology, Cambridge, MA~02139, USA \\
$^{2}$ Department of Physics, New Mexico State University,
Las Cruces, NM~88003, USA \\
$^{3}$ Department of Physics, Florida Atlantic University,
Boca Raton, FL~33431, USA}
\begin{document}

\date{Accepted MNRAS 2015 January 23}

\pagerange{\pageref{firstpage}--\pageref{lastpage}} \pubyear{2014} 
\maketitle

\label{firstpage}

\begin{abstract}
A formula is derived for the combined motional and gravitational Doppler
effect in general 
stationary axisymmetric metrics for a photon
emitted parallel or antiparallel to the assumed circular orbital motion
of its source. The same formula is derived by both the eikonal approximation and
Killing vector approaches to elucidate connections between observational astronomy and 
modern Relativity. The formula yields expected results in the
limits of a moving or stationary source in the exterior Kerr and
Schwarzschild metrics   and is useful for 
broad range astrophysical   analyses.
\end{abstract}


\section[]{Introduction}
Light is the predominant means of detection and observation on
astrophysical scales. However, light undergoes distortion  
due to various effects 
before being received on or near earth.
The chief phenomenon in this regard is the Doppler shift
in the observed frequency of a photon.  We use the term ``Doppler shift'' in a generalized sense throughout this paper, to refer to any effect that causes a photon's frequency at detection to differ from that which it had when emitted.  Thus, we are interested in a combination of factors.
On the one hand, there is the kinematic effect resulting from
relative motion of emitter and observer.  This special relativistic effect, which commonly is called the ``Doppler shift,'' is analyzed using the Minkowski metrics on the tangent spaces at the spacetime events where a given photon is emitted and received.  On the other hand, there are general relativistic effects due to the photon's propagation through the intervening region of curved spacetime.  These are usually called ``gravitational redshifts.''  They
include not only the effect of the 
Coulombic
potential well implied
by the mass distribution, but also the frame dragging effect 
caused by its spinning motion.
%
%
Experimentally, of course, only a combination of these effects can be observed.  We treat them all together under the name ``Doppler shift'' accordingly.

 
The Kerr--Doppler effect resulting  from the combination of effects listed above has been given in \citet{fanton} for the Kerr black hole geometry.  We have two aims in the present paper.  The first is to extend the analysis to arbitrary Kerr-type (\textit{i.e.}, stationary and axisymmetric) spacetimes,  e.g., the coarse-grained, diffuse internal spacetime of a spiral galaxy.  The extension allows  new, high resolution   observations \citep{EventHorizon}  to distinguish between 
different model based predictions.  
The second is to illustrate full derivations, in parallel formalisms used in the astrophysics   and relativity communities,  to draw attention to assumptions involved at each step in the analyses and to illustrate the utility of each approach.   The astrophysics derivation is based on coordinates and an effective optical index of refraction for wave propagation~\citep{1996Narayan}, and the modern relativity construction  on conserved quantities along particle geodesics and Killing vectors~\citep{Wald}.  Naturally, we will see that the two lead to the same final expressions.

Various  frequency-shift  effects have to be taken into account in a wide array of
astrophysical contexts.  These include, for example, 
the interpretation of X-ray
spectral distributions originating from black hole accretion disks
\citep{asaoka,bromley,cunning,fabian,fanton,laor,li,martocc}, 
and the implementation of satellite navigational systems
\citep{bahder,teyssan,pasc}.
Depending on the specific context, emphasis has to be placed on
different aspects influencing the 
shift.  For example, where
satellite navigational systems are concerned, the gravitational fields
near earth are sufficiently weak to allow for the perturbative
treatment of general relativistic effects.  However, specific
details of the earth's gravitational field, such as deviation from
exact spherical shape, are important. Thus, while \citet{teyssan}, using a 
linearized axi-symmetric metric,  expands the
Doppler shift in $1/c$ to fourth order, it includes not only its dependence
on the earth's mass and angular momentum, but also the quadrupole
moment of the mass distribution. In the context of black holes, the
underlying metric has fewer complicating features, but one cannot
invoke a perturbative expansion in $1/c$, nor may one always neglect
deviations of light geodesics from straight lines. Therefore, in general
one is forced to integrate the geodesic equations numerically,
cf.~\citet{asaoka,fanton,laor,li}.

The present note is intended for a certain special class of geometries, wherein simple analytical
results for the Doppler shift
can be obtained in order to  give guidance for    more 
complex geometries or to distinguish between different model based assumptions. One such example is treated in \citet{radoszpl},
analyzing   situations in which the Doppler shift factorizes
into the kinematic contribution and the general relativistic contribution
(in general, these contributions are intricately entangled). Another example where 
  the formula derived in this paper is  useful is in the analysis presented by \citet{Schonenbach}, 
contrasting the predictions of emission from the accretion disks of  central galactic nuclei in  General Relativity versus pseudo-complex General Relativity.

The geometries under
consideration here comprise all stationary axisymmetric metrics of the Kerr
type, i.e., all metrics independent of time $t$ and azimuthal angle
$\varphi $ in polar (Boyer-Lindquist) coordinates $(t,r ,\theta, \varphi )$,
with $g_{t\varphi } = g_{\varphi t} $ the only nonvanishing off-diagonal
elements. In such metrics, we consider a test particle moving in the
$\varphi $-direction (which is the case for emitters in circular orbits
in the equatorial plane, and also for emitters at the apsides of other
orbits in that plane), and emitting in (or against) that same direction.
In this setting, the Doppler shift observed by a receiver in asymptotic
flat space can be given without recourse to a perturbative expansion.
The treatment does not yield information about the light geodesic; thus,
while the result for the Doppler shift derived here in itself is exact,
its practical application will usually require complementary 
information about, say, the position of the emitter. E.g., if one is interested in
reconstructing a radial velocity distribution from the Doppler-shifted light
observed, one has to link the Doppler shift to the radius of the emitter's
orbit, which in general may require a numerical treatment of the deflection
of the light ray, as mentioned above. On the other hand, due to
the general form of the class of metrics considered, the Doppler formula
given here is expected to be of use in the context of interior Kerr-type
metrics describing space-time structure close to, or inside, extended
rotating matter distributions.
Whereas the case which we present here may require further generalization
depending on the context of application, we nonetheless regard it as
representative of what we would have to consider, e.g., to obtain a sense
of the scale of the observational signature of frame dragging effects for
orbital telescopes.

The Doppler shift in the geometries described above will be derived using
two different methods, namely, by an eikonal approach in section \ref{eiksec},
and by employing invariants constructed from the Killing vectors of the metric
in section \ref{killsec}. The special case of the Kerr metric
\citep{chandra,oneill} and further limiting cases are considered
at the end of section \ref{killsec}.

 \section[]
{Eikonal Approach}
\label{eiksec}
This approach arises from two aspects of the physics. The first is
that   we may define the    energy $E$, and thus the frequency $\omega$,  of a photon  as measured by any observer
at any location by \citet{Hartle}  
as
\begin{equation}
\omega_o= -{\bf u}\cdot {\bf k} =  -u^\alpha k_\alpha ,
\label{eq:willtest}
\end{equation}
 
where ${\bf u}$ is the 4-velocity of the local observer 
and ${\bf k}$ is the momentum $4-$vector of the photon at the observer's location (natural units). 

The second aspect is that, in the approximation we are using,  
 the photon can be propagated out to asymptotic
infinity, enduring only negligible bending of its ray path. This argument
allows us to use the eikonal approximation to  
read off an effective index of refraction from the wave equation.
We assume that the scalar
wave equation will capture the relevant physics in the eikonal limit.
This wave equation for a wavefunction $\Psi(x)$ is
\begin{equation}
\Box \Psi  =  \frac{1}{\sqrt{g.}} \frac{\partial}{\partial x^\mu}
\left( g^{\mu\nu} \sqrt{g.} \frac{\partial}{\partial x^\nu} \right)\Psi
=0 \ ,
\label{eq:extrn}
\end{equation}
where $g^{\mu\nu}$are the contravariant metric components,
and $(g_.)$ is the determinant of the matrix of covariant components
$g_{\mu\nu}$.  We consider general Kerr-type metrics in Boyer-Lindquist 
coordinates, $(t,r, \theta, \varphi)$, whereby the nonzero $g_{\mu\nu}$ are
$g_{tt}, g_{t\varphi},g_{\varphi \varphi }, g_{rr },$ and $g_{\theta
\theta} $, independent of $\varphi$ and $t$.  

We write an eikonal approximation for the wave function,
\begin{equation}
\Psi({\bf r}) =\Psi_o \exp(i \phi) ,
\label{eq:3}\end{equation}
  where $ \Psi_o$ is a constant amplitude and $\phi$ is the integral along the photon path of the differential phase

\begin{equation}
d\phi =k_\alpha dx^\alpha.\label{eq:4}
\end{equation}
We consider a photon emitted in a direction tangent to a circular orbit, so that in the 
neighborhood of the emission point, $k_r=k_\theta=0$.  We also write
\begin{equation}
k_\varphi= {\bf e}_{\varphi}\cdot {\bf k}=k\sqrt{g_{\varphi\varphi}},\label{eq:5}
\end{equation}
where we have defined the wavenumber $k=\hat{\bf e}_{\varphi}\cdot   k=k_\varphi /\sqrt{g_{\varphi\varphi}}$.  Here $\hat{\bf e}_{\varphi}={\bf e}_{\varphi}/\sqrt{g_{\varphi\varphi}}$ is simply a unit vector parallel to the covariant
basis vector ${\bf e}_{\varphi}$; it is \emph{not} part of an orthonormal tetrad.  (Note that while $k_\varphi$ is dimensionless, $k$ has dimension of inverse length, appropriate for a wave number.)  Also, using eqs. (\ref{eq:3}), (\ref{eq:4}) and  (\ref{eq:5}), and the fact that the metric is independent of $t$, the wave function in a neighborhood of the photon  emission point $(r, \theta, \varphi=\varphi_o)$ is
  
\begin{equation}
\Psi(x) =   \Psi_o \exp(-i \omega t) \exp\left(ik\sqrt{g_{\varphi \varphi}}
(\varphi-\varphi_o) \right)
\label{eq:wakey}
\end{equation}
where $\omega=-k_t >0$ is the invariant  angular frequency of the wave with respect 
to the time coordinate.   It is the frequency measured by an observer at rest at spatial infinity, where $k^t=-k_t$.

Inserting eq.~(\ref{eq:wakey}) into eq.~(\ref{eq:extrn}) yields
\begin{equation}
\omega^2 [-g^{tt} + 2ng^{t\varphi} \sqrt{g_{\varphi \varphi } } -
n^2 g^{\varphi \varphi } g_{\varphi \varphi } ] \Psi = 0.
\label{eq:vhc}
\end{equation}
where the effective index of refraction $n$ is defined by
\begin{equation}
v_{photon}=\omega/k=c/n
\label{eq:ls}
\end{equation}
and $v_{photon} $ is the coordinate light speed at the point of emission.
The general solution of eq.~(\ref{eq:vhc}) is
\begin{equation}
n = \frac{ -g_{t\varphi} -  \sqrt{(g_{t\varphi})^2-
(g_{\varphi\varphi})(g_{tt})}}{ g_{tt}\sqrt{g_{\varphi\varphi}} }
\label{eq:index_yayaya}
\end{equation}
Note that the denominator is negative,
since $g_{tt} \approx -1 + 2\Phi/c^2$ in weak-field Kerr-type metrics,
where $\Phi$ is the Newtonian gravity potential~\citep{Hartle}. Therefore we
chose the $(-)$ sign preceding the square root in order to obtain
a positive $n$.
 
For an orbiting observer, the spatial part of the product in 
 eq.~(\ref{eq:willtest}) is 
  
\begin{equation}
 k_\varphi u^ \varphi 
= k\sqrt{g_{\varphi\varphi}} \ \Omega u^t
\label{eq:happybaby}
\end{equation}
where we have defined the angular velocity
\begin{equation}
\Omega = d\varphi/dt = \frac{u^{\varphi}}{u^t}
\label{eq:curious}
\end{equation}
which is a constant for a circular orbit. Note that i) 
$\Omega$ may be negative or positive, for source motion away from or
towards the asymptotic observer, respectively; and ii) for 
sources in circular orbits the magnitudes of these two angular 
velocities will be different in Kerr-type metrics because of 
their nonzero $g_{t\varphi}$.

From eqs.~(\ref{eq:ls}), (\ref{eq:happybaby}), and (\ref{eq:curious}), 
we obtain
\begin{equation}
\omega_o
= \omega u^t \left[1 -n \Omega \sqrt{g_{\varphi\varphi}}  \right].
\label{eq:kerrdopl5}
\end{equation}
We obtain an expression for
$u^t$ from the constraint ${\bf u}\cdot{\bf u}=-1$:
\begin{equation}
u^t = 1/\sqrt{\left(-g_{tt}-2\Omega g_{t \varphi} -\Omega^2
g_{\varphi\varphi}\right) }.
\label{eq:sippycup}
\end{equation}
Then, substituting
eq.~(\ref{eq:curious}) and (\ref{eq:index_yayaya}) into
eq.~(\ref{eq:sippycup}) yields the general Kerr Doppler formula for a
source moving directly toward or directly away from the asymptotic observer:

\begin{equation}
\frac{\omega_o}{\omega} =
\left[\frac{ g_{tt}+\Omega \left( g_{t\varphi} +\sqrt{(g_{t\varphi})^2-
g_{\varphi\varphi}g_{tt}}\right)}{g_{tt}\sqrt{-g_{tt}-2\Omega
g_{t\varphi}  -\Omega^2g_{\varphi\varphi} } }  \right].
\label{eq:kerrdopl}
\end{equation}
 
\section[]
{Killing Vector Approach}
\label{killsec}
This derivation will take advantage of  the conserved quantities of time-like and null geodesic 
motion which  result from  the    one-parameter families of symmetries of the space-time, described 
by the Killing  vector fields  \cite[(C.3.1)]{Wald}.   For stable, circular, equatorial orbits that most
closely approximate those in spiral galaxies \citep{Klypin}, the relevant KV are the time-like  ${ \bf \xi} = (1,0,0,0)$ and the axial  $ {\bf \eta} = (0,0,0, 1)$, for Kerr-type metrics in Boyer--Lindquist-type coordinates $(t, r,  \theta, \varphi )$.
The norms of these Killing vector fields are defined  as is done in \cite{Wald}. 

 The 4-velocities  for  such  particles, ${\bf u}=(u^t,0, 0,  u^{\varphi} )$,  can be expressed  in terms of  the KV as:
\begin{equation}
{\bf u }  =A {\bf \xi } +B {\bf \eta }
\label{eq:suchtat}
\end{equation}
for  $A$ and $B$  coefficients   determined by the symmetries which give the 
  conserved  
``energy'' $E$ and ``angular momentum'' $L$ (both per unit mass)   of the particles. The conserved quantities  
 $E = -{\bf u}\cdot  { \bf \xi}$ and 	$L = {\bf u}\cdot  { \bf \eta}$ are defined as   in  \citet[eq.~(15.17)]{Hartle}.  
 
  By inspection, $E$ and $L$  can be combined into the system
of equations:
\begin{equation}
\begin{pmatrix} -E \\ L\end{pmatrix} =
\begin{pmatrix}
g_{tt}  &  g_{t\varphi}  \\
g_{t\varphi}   &  g_{\varphi \varphi}
\end{pmatrix}
\begin{pmatrix} A \\ B \end{pmatrix}.
\label{KVmatrix}
\end{equation}
Matrix inversion  then  yields $A$ and $B$ in terms of $E$ and $L$, such that eq.~(\ref{eq:suchtat}) can be written as:
\begin{equation}
 {\bf u} = A {\bf \xi} +B {\bf \eta}
= \frac{ g_{\varphi \varphi}E  + g_{t \varphi}L}{\kappa} {\bf \xi}
-\frac{g_{t \varphi}E  + g_{t t}L}{\kappa}  {\bf \eta} \ 
\label{eq:MTWfour}
\end{equation}
  for $\kappa = (g_{t\varphi}^2 - g_{tt}g_{\varphi \varphi})$.\\
  
For  time-like geodesic motion, the condition ${\bf u \cdot u}=-1$ yields:
\begin{equation}
-1= \frac{ g_{t t}L^2 + 2g_{t \varphi}EL
+ g_{\varphi \varphi}E^2}{\kappa} \ .
\end{equation}

Similarly for the  line-of-sight  photons  emitted from such particles, the  initial 4-momentum is  in only the  two KV  directions $\xi$  and $\eta$.  Thus the photon 4-momentum ${\bf k}$ can be expressed  exactly as is ${\bf u}$ in  eq.~(\ref{eq:MTWfour}), but  replacing the conserved particle quantities $E$ and $L$ with those of the photon
  $e$ and $l$.   The photon null condition $0=k^a k_a$ then  yields a relations between $e$ and $l$:
\begin{equation}
\frac{l}{e}
= \frac{-g_{t\varphi}- \sqrt{\kappa} }{g_{tt}}=n\sqrt{g_{\varphi\varphi}},
\label{eq:exis}
\end{equation}
which connects  the ratio $l/e$ to the previously obtained refractive index $n$ of
eq.~(\ref{eq:index_yayaya}).    Again,    the positive root is used   because our photon moves in the forward direction.

Then, the   inner product of the particle   and  photon  4-vectors  can be expressed as
\begin{equation}
\omega_o =  -{\bf u}\cdot {\bf k} = e\frac{ g_{\varphi \varphi} E + g_{t
\varphi}(E\frac{l}{e} +L) + g_{tt}L\frac{l}{e}}{\kappa} .
\label{eq:shammy}
\end{equation}

Finally, given that  ${\bf k}\cdot  {\bf \xi}=\mbox{const.} $
along the photon path \citep[(C.3.1)]{Wald},   the stationary observer at infinity 
whose   four-velocity  is ${\bf u}_\infty = {\bf \xi}$ measures the received photon frequency  as
$\omega= {\bf u}_\infty \cdot {\bf k}  = e \ $.
   This  yields from eq.~(\ref{eq:shammy}) 
 the ratio:  
 \begin{equation}
\frac{\omega_o}{ \omega } = \frac{1}{\kappa}
\left[  g_{\varphi \varphi} E + g_{t \varphi}(En \sqrt{g_{\varphi
\varphi}} +L)
+ Ln g_{tt}\sqrt{g_{\varphi \varphi }}
\right].
\label{eq:previous}
\end{equation}
Substitutions for $E$ and $L$, from \citet[eq.~(15.17)]{Hartle},  into 
  eq.~(\ref{eq:previous}) then gives
\begin{equation}
\frac{\omega_o}{\omega } =
u^t\left(1- \Omega n \sqrt{g_{ \varphi \varphi}}  \right),
\end{equation}
for $\Omega$ as in eq.~(\ref{eq:curious}). This is the same result 
obtained by the eikonal
approach in  eq.~(\ref{eq:kerrdopl5}).   

Note,  the 
same result can be derived more quickly   in the special case when the relationship 
between $\Omega$ and $ r$  is known, such as when Keplerian orbits become physically reasonable far from the accretion disk or  when  a normalization condition is applied which reduces   eq.~(\ref{eq:willtest})  to 
  $\omega_0=-{\bf u}\cdot {\bf k}= u^t  k_t-u^{\phi}  k_{\phi}= u^t e-u^{\phi}  l$.  Used together with the null  condition  for the wave vector,  the result in eq.~(\ref{eq:kerrdopl5}) can be derived in three lines.  However, the   relationship between $k_t$ and $k_\varphi$, and the conserved quantities $l$ and $e$,  is in general  more complicated  \cite[see eq.~(15.7)]{Hartle}.  In the general case, the form derived in   eq.~(\ref{eq:previous}) in arguably more useful,  where the relationship between the orbital frequency $\Omega$ and the $r$ coordinate is  most readily expressed  by the relationship between $E$, $L$, and $r$.   These are the correct set of variables that you most easily calculate with in General Relativity.

In the special case of the Kerr metric proper \citep{chandra,oneill},
\begin{align}
g_{tt} &= -(1-2Mr/\Sigma )  , \notag\ \  
g_{\varphi \varphi } &= ((r^2 +a^2 )^2 - a^2 \Delta \sin^{2} \theta )
\frac{\sin^{2} \theta }{\Sigma}
 \label{coeffic}
\end{align}

\begin{equation}
g_{\theta \theta } = \Sigma \ , \ \ \ \
g_{rr} = \Sigma / \Delta \ , \ \ \ \
g_{t\varphi } = g_{\varphi t} = -2M a r\frac{\sin^{2} \theta }{\Sigma}
\end{equation}
where
\begin{equation}
\Sigma = r^2 + a^2 \cos^{2} \theta \ , \ \ \ \
\Delta = r^2 + a^2 - 2Mr \ ,\notag
\end{equation}
and for an emitter in the equatorial plane, $\theta = \pi /2$, we have
verified that our Doppler formula (\ref{eq:kerrdopl}) coincides with the
expression given in \citet{fanton}, cf.~also \citet{asaoka,cunning,li}.
Specializing further to the Schwarzschild metric limit by setting $a=0$,
with $M=\frac{M_s}{c^2}$ for a source of mass $M_s$, one obtains
from eqs.~(\ref{eq:index_yayaya}) and (\ref{eq:sippycup}),
again for an emitter in the equatorial plane,

\begin{align}
n = & (-g_{tt})^{-1/2}  =  \left( 1-\frac{2M}{r} \right)^{-1/2}  \notag\\  
u^t =& c\left( 1-\frac{2M}{r} -\frac{\Omega^2r^2}{c^2} \right)^{-1/2}, \label{ecku}\\
\end{align}

where the appropriate factors of $c$ have been inserted. Then after a little
algebra, eq.~(\ref{eq:kerrdopl}) yields 
\begin{equation}
\frac{\omega}{\omega_o} =\left[ 
\left(1-2M/r\right)  
\left(\frac{ (1-2M/r)^{1/2}  +\Omega r/c}{(1-2M/r)^{1/2}-\Omega r/c}
 \right)
\right]^{1/2}.
\label{eq:ifandso}
\end{equation}

Clearly, this result has the correct limits: One obtains the usual
gravitational redshift for non-moving optical sources ($\Omega =0$),
and the usual longitudinal  Lorentz Doppler ratio for $M=0$ but $\Omega
r/c = v/c$, where the relative source-observer velocity $v\approx \Omega
r$ can be positive or negative. Note that, for a source in circular orbit,
$M/r\approx v^2/c^2$, so eq.~(\ref{eq:ifandso}) yields the usual
Lorentz Doppler formula to first order in $v/c$.

\section[]{Summary and Discussion}

In this paper, we derived a formula for the motional and gravitational
``Kerr--Doppler'' effect for a photon emitted tangentially to the motion of
its source. The source was restricted to be moving azimuthally in any
Kerr-type metric, i.e., any metric which when expressed in polar coordinates
$(t,r ,\theta, \varphi )$
has only $g_{tt} $, $g_{rr} $,
$g_{\varphi \varphi } $, $g_{t\varphi } $, and $g_{\theta \theta } $ nonzero
and functions only of $r$ and   $\theta $. The formula, eq.~(\ref{eq:kerrdopl}),
provides the frequency of the photon measured by an observer at rest at
spatial infinity in terms of that measured by a local observer co-moving with
the source. We showed that the formula yields expected results in the limits
of an orbiting or stationary source in the exterior Schwarzschild metric and
a moving source in flat space, and also agrees with the result of
\citet{fanton} for the exterior Kerr metric.

In obtaining the formula, we utilized two seemingly different approaches,
an eikonal approximation solution to a scalar wave equation in the
Kerr-type metric, and a Killing vector (KV) representation of both the
source circular motion and the photon motion. The two approaches produced
the same formula, because despite apparent dissimilarities the underlying 
  physics is the same. For example, the local propagation, or wave 3-vector used in
the eikonal approach is (proportional to) the local photon 3-momentum used
in the KV approach. While the KV approach is limited to the particular
highly symmetric application that we treated, that of a photon emitted
tangentially to the circular orbit of a source, it allows analysis in a more modern
relativistic context,  in which the  relationship  between 
the orbital velocity, $\Omega$, and the radius is determined by the   the conserved quantities
of the Lagrangian, 
$E$ and $L$. On the other hand, the eikonal method should
be applicable for a local photon propagation $3$-vector in any direction
relative to the source motion, cf.~\citet{fanton} for the special case of
the exterior Kerr metric. The wave equation (\ref{eq:extrn}) would
then yield a different expression for the local effective refractive index
than we obtained in eq.~(\ref{eq:index_yayaya}) for tangential emission.

The categorical derivation of the formula from the two perspectives given here  allows construction of model based  predictions,
such as those made in   \citet{Schonenbach},  which can then be used to distinguish feature of the models' physics. In example, 
the predictions made in  \citet{Schonenbach} can 
 distinguish between  General Relativity and pseudo-complex General Relativity (pcGR) in late 2015
 when the   Event Horizon Telescope(EHT) \citep{2011Doelman, 2014Fish}  begins to collect data 
 at our Galaxy's nucleus and that of   M87.   The advent of such high resolution observations, 
 where we can distinguish fine features in emission lines from accretion disks, makes 
 the formula given here important as we move forward in our analysis of the cosmological 
 gravity theory.

 \section{acknowledgements}The authors wish to acknowledge the  referee for   
substantial comments which lead to  clarifications in the presentation of the derivations and applicability of the Kerr-Doppler formula,  and demonstrated
an  important normalization condition which can simplify the results   tremendously under specific circumstances.  S.C. wishes to thank the Massachusetts Institute of Technology Martin Luther King Scholars Program for the opportunity to work in a  vibrant community.

 \bibliography{LCM}{}
\bibliographystyle{mn2e}
 \bsp
 
\label{lastpage}
\end{document}